\documentclass[11pt]{article}
\usepackage[a4paper, total={16cm, 23.5cm}]{geometry}
\usepackage{graphicx}
\usepackage{hyperref}
\usepackage{xcolor}

\begin{document}

%\RUNTITLE{Graph theory: original teaching supports}

\title{Innovative ideas for teaching supports: \\
Application to Graph theory}

\author{Nicolas Catusse, Hadrien Cambazard, Nadia Brauner, Bernard
Penz, Florian Fontan\\
Univ.  Grenoble Alpes, CNRS, Grenoble INP,
G-SCOP, F-38000 Grenoble, France}

\maketitle

\noindent {\bf Abstract} Teaching graph theory with the most adequate
tools requires time and ideas. We present how an open community of
teachers shares contents and ideas on an innovative platform. The
objective is to get the students autonomous in their training with
activities that give them immediate feedback on their understanding.
Beyond learning, the very large collection of exercises of various
levels can also be used to evaluate the student's level. The proposed
activities can be algorithm's code in classical programming languages
(e.g. Java, Python) that the student can test with predefined tests
proposed by the teacher or collections of generated questions. \\


\tableofcontents

\ \\[1cm]

%\com{motivation au demarrage de la plateforme}
This article presents innovative ideas for teaching graph theory, and
how they have been implemented in a new pedagogical platform:
caseine.org.  Graph theory is taught at university at different levels
(bachelor, master) and for different audiences (computer science,
mathematics, management, engineering, etc.).  Graph theory teaching
can be considered from various perspectives: practical problem
modelling, graph culture, algorithmic analysis and programming,
mathematical proofs, etc.  Designing activities that allow students to
practice these skills or to assess their understanding automatically
requires various types of advanced learning mechanisms, novel
ideas, and might be time consuming.  An open community of teachers has
therefore decided to pool their experiences, their course contents,
and to develop new activities together.  Providing students with a
variety of contents, self-assessment tests, auto-evaluating
programming exercises allows them to have a better feedback on their
works thus increasing their autonomy.  The consequence is that the
teacher, being delivered from basic tasks, is better available to
guide the students in their progress.

\begin{figure}
	\center \fbox{\includegraphics[width=0.9\textwidth]{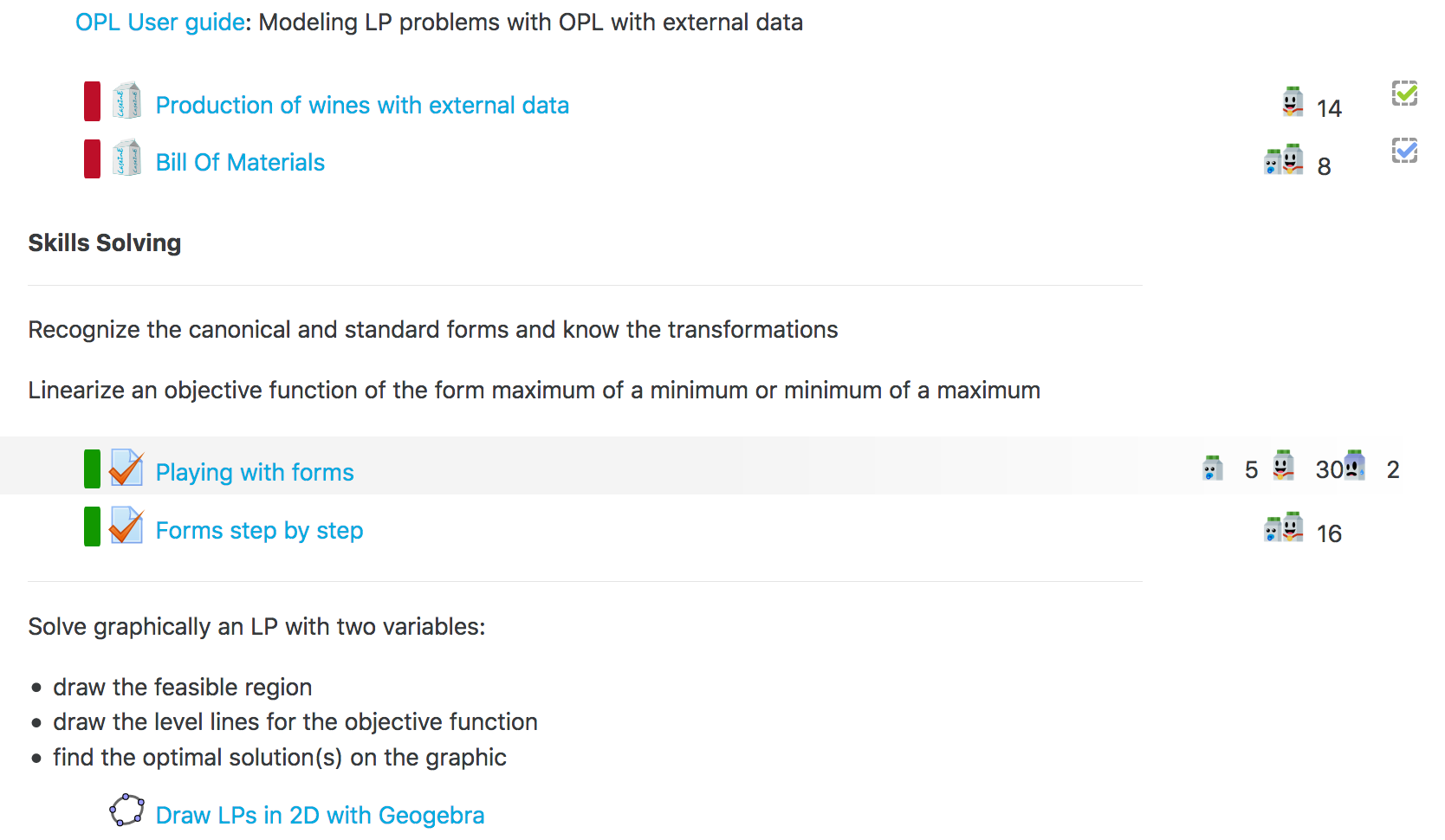}}
	\caption{A
	screenshot of the student view of a caseine course.}
	\label{fig:caseine}
\end{figure}

%\com{generalites sur Caseine}
Those ideas and activities have been implemented on the caseine
platform\footnote{see a 3-minute tour of the platform 
\href{https://moodle.caseine.org/mod/page/view.php?id=19571}{caseine.org}}. It is an open learning platform dedicated to
university and secondary school teaching in Computer Science and
Mathematics, including Operations Research (OR) education.  On
caseine, an international community of OR teachers shares ideas, a
wide variety of pedagogical contents and advanced tools for
evaluation.  Anybody with an academic account can access open courses
as a student.  Non-academics can ask for an account.  A professor can
create courses and use the learning tools and shared contents.
Caseine is based on the famous and widely used Learning Management
System Moodle which offers students a learning environment and allows
to monitor students' progress (see e.g. Figure~\ref{fig:caseine} where
{\em Bill of Materials} is a programming lab, {\em Playing with forms}
is a quiz...).  The originality of caseine.org, compared to a 
classical university Moodle instance, comes from the fact that it is an open platform offering at the same time (1) a pedagogical environment for university training where the teachers can build their own course and (2) coding exercises with automatic and self-evaluation and (3) all sorts of pedagogical activities shared by an open community of teachers. 

%\com{plan du papier}
This paper presents activities and ideas developed by
the teachers for graph theory training using the original tools of the
caseine platform.  In Section~\ref{sec:AutoLP}, we detail the main
contribution of the platform, namely the activities for graph
algorithm learning.  In Section~\ref{sec:question_bank}, we present
the participative advanced question banks, from basic true-false
questions to more sophisticated questions and game activities.  In
Section~\ref{sec:caseine}, we present the platform, its architecture,
and the benefits encountered when using it.

\section{Graph algorithms' implementation learning}\label{sec:AutoLP}

%{\color{red} 
A graph theory course often requires the programming
of the main algorithms so that the students understand them better.
Indeed, succeeding in programming an algorithms can be an effective
complement for understanding it deeply than only a paper
implementation on some examples. It also allows the students to work
actively on the useful data structures and the algorithmic complexity
of the solutions considered.

We will present how automatic evaluation can be helpful in this 
context. %
%}
Automatic evaluation consists in offering automatically (without
teacher intervention) an evaluation with possibly some customized
comments on the answer given by the student to a question.  Of course,
caseine platform offers classical automatic evaluation tools like
multiple-choice questionnaires, questions with calculated answer, etc. (see Section~\ref{sec:question_bank}).
However, it goes further by offering tools to evaluate student's
programming code in several programming languages (Java, Python, C...)
or student's mathematical models.  The latter case can be the 
modeling of well-structured problems \emph{i.e}  with a formal language and well defined statements, like linear
programs. Then, the evaluation consists in verifying that the underlying
mathematical object (a polyhedron for linear programs) is indeed the
expected one, without doing too restrictive hypothesis on the
student's model (see  \cite{Cambazard21} for more details).  In this section, we present programming exercises
with automatic evaluation (we call them {\em labs}) on classical graph
algorithms with adequate data structures.

\subsection{Auto-evaluation coding tools: on the student's side}

The programming activities described in this section are intended for students at a bachelor 
level in Computer Science and Applied Mathematics training. The objective is 
to understand graph implementations and to program classical 
algorithms with the adequate data structure.

The two interfaces we present (web navigator or IDE tool)  offer the following 
features for the students:
\begin{itemize}
	\item push the proposed code onto the platform to submit it to
	the teacher;

	\item run the tests programmed by the teacher and see the
	results to validate functionally the proposed code;
	
	\item potentially ask for teacher comments and get them;
	
	\item possibly get a mark for their work.
\end{itemize}

\begin{figure}
	\center \fbox{\includegraphics[width=0.95\textwidth]{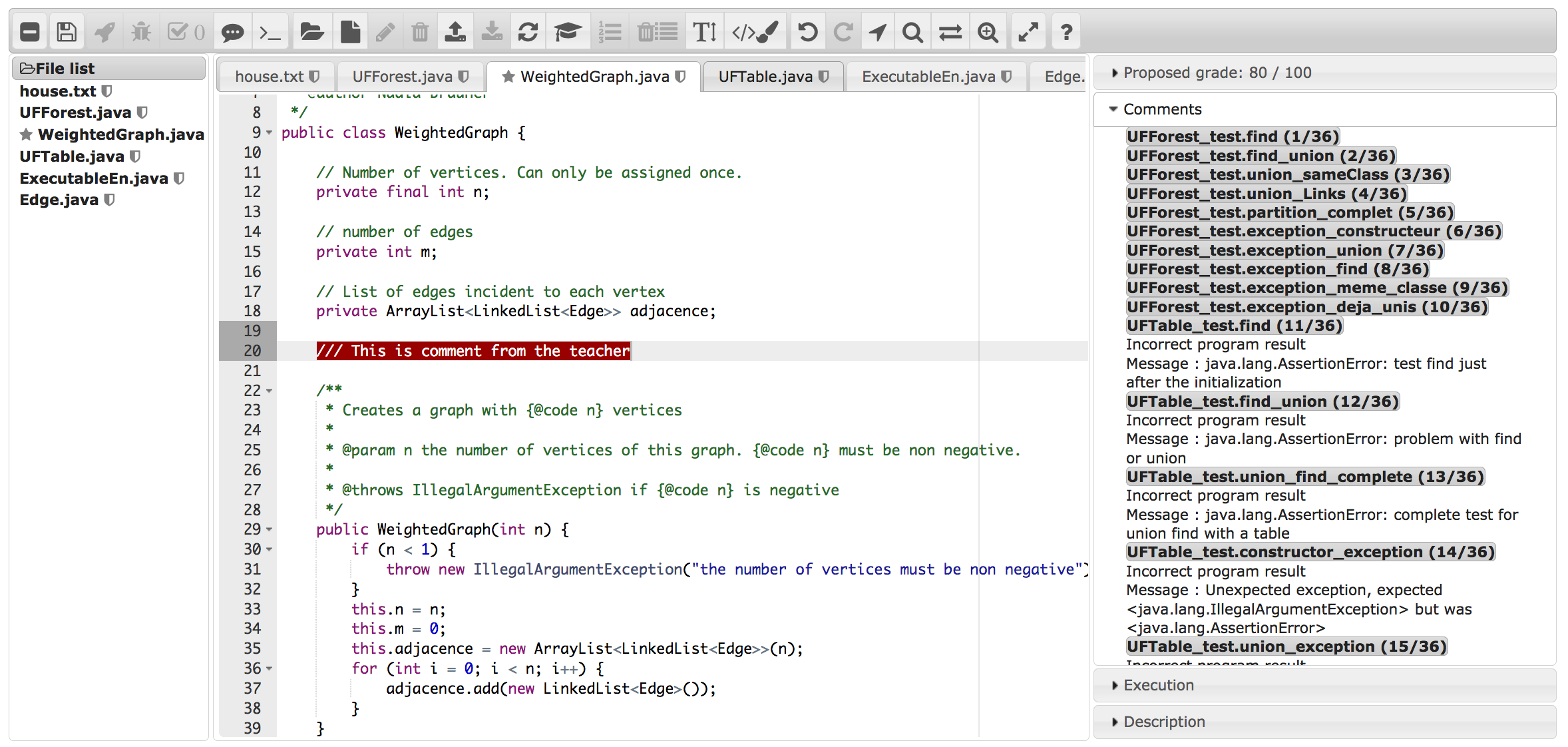}}
	\caption{Programming interface within the Web navigator.} 
	\label{fig:VPL}
\end{figure}

The first interface is based on the VPL Moodle 
plugin\footnote{\href{https://vpl.dis.ulpgc.es/}{Virtual Programming Lab} developed by Juan Carlos 
Rodríguez-del-Pino.}. This plugin has been slightly adapted for our 
usage. Figure~\ref{fig:VPL} presents the plugin interface. The code 
can be edited in the middle part of the window. The 
tick-mark button allows the students to evaluate their code by 
running the tests settled by the teacher. This button also saves
the code within  the system allowing the teacher to see and comment 
it. The right part of the 
window presents the results of the tests together with helping 
comments and a proposed mark. 

\begin{figure}
	\center \fbox{\includegraphics[width=0.95\textwidth]{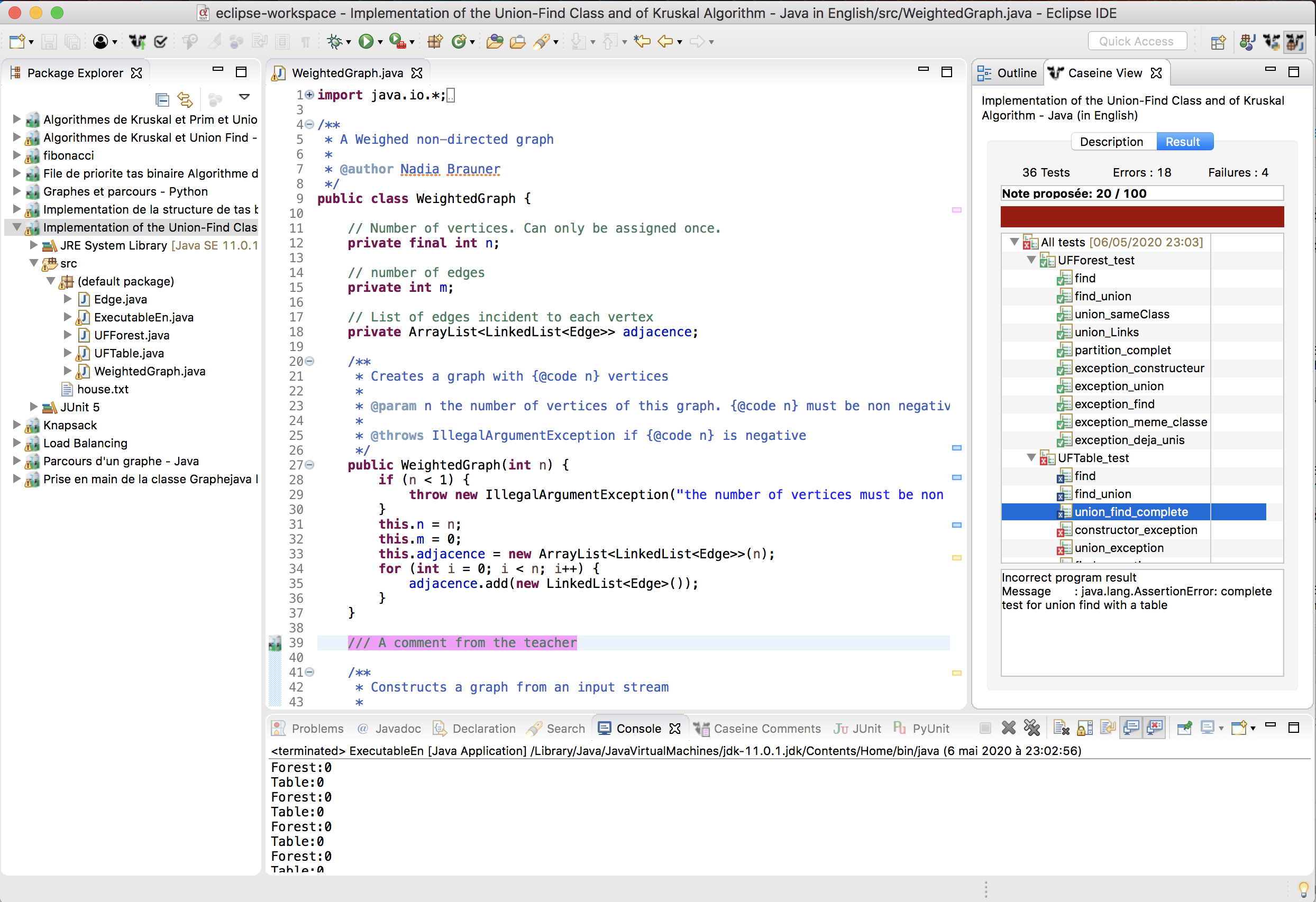}}
	\caption{Caseine View and tools within the Eclipse IDE.} 
	\label{fig:eclipse}
\end{figure}

This web interface has the huge advantage to be available on any
machine, having no installation needed.  However, for programs with a
large number of lines, this interface is not rich enough.  Therefore, 
students are encouraged to work within an external IDE  (Integrated Development Environment) with a caseine perspective 
that allows to pull code from the platform, push code, launch the 
tests and get the results as well as the comments of the teachers. Figure~\ref{fig:eclipse} shows the caseine 
view within an Eclipse Workspace\footnote{\url{https://www.eclipse.org/}}. A similar plugin is available for 
Visual Studio Code and VSCodium\footnote{VPL Client Extension for 
VSCode by Guillaume Huard, available on the
\href{https://marketplace.visualstudio.com/items?itemName=perelabat.vpl}{VSCode Marketplace}} and for IntelliJ\footnote{Caseine VPL plugin for IntelliJ available on the \href{https://plugins.jetbrains.com/plugin/19472-caseine-vpl}{JetBrains Markeplace}}. 

The teachers can have access to all submissions via the web interface. They can see the code, comment it and give access to the students 
to their comments, get the automatic mark and, if necessary, adjust it. 

\subsection{Creating a lab: on the teacher's side}\label{sec:GenTests}

Various tools have been developed to help teachers creating 
programming labs. The graph labs have been developed in Java and 
Python. Therefore, we detail the tools offered in those two 
languages. However, the labs can be developed in any programming 
language with more or less effort depending on the already available 
tools for the language.

To develop a programming lab, the teacher has to propose a basic 
code to be given to the students (a code's skeleton), an expected/corrected code that will be 
available for the teachers of the course and that might be used by 
the tests, and a collection of tests (see section~\ref{sec:tests}). In both languages, the teacher 
writes the complete expected code and puts annotations to indicate 
where the code should be replaced by a TODO comment. In Java, the 
tests are written in the classical unit testing framework 
JUnit\footnote{\url{https://junit.org/}}. In 
Python, they can be written in Unittest (also referred to as PyUnit) 
or in a small  dedicated language. We also use Doctest to let the 
students know what is expected and test their code. A message can be 
added to each test to give students indications on their code. 

The adequate files (for the students, for the teacher, test files) 
are then generated and downloaded into the platform to be available 
from the various interfaces.

\subsection{Generating tests}\label{sec:tests}

The objective of the tests are to settle whether the student's code
answered the question.  In the case of Graph algorithms, we want to
verify that the code indeed calculates the proper object (functional
testing).  For instance, consider the case of Kruskal's Algorithm for
minimum spanning trees. 

Some tests are dedicated to 
verifying gradually and manually the results: we give the students toy
examples with the expected value.  The first test is on a simple
connected graph with unitary edge.  The system indicates the student
whether the program indeed returns a tree.  Other unitary graphs are
also proposed for testing with various sizes, some of them are
available for the students together with the expected value, others
are hidden.  Once the students know that the return object is 
correct (a spanning tree in this case), they can concentrate on the weighted 
case and verify
that the tree returned by the algorithm has the correct
weight.  Some tests check the limit cases (e.g. non-connected graphs,
stable graphs) or board effects (e.g. the graph is not modified by the
students code).  This activity also contains functional tests on the
union-find structures.

The students, being able to see some of the tests proposed by the 
teacher, can improve their own testing skills.

The automatic tests allow to gain time for the teacher and to have 
immediate feedback for the student but they do not exempt the teacher 
from some proofreading of the code quality for instance. Reading the student's code also allows to add tests on 
special cases which might be not covered.

\subsection{A collection of shared labs}

All programming labs are based on a Graph class that offers basic
methods to handle graphs: create a graph, read a graph from a file,
get the neighbors of a vertex...  We now describe some available
programming exercises.

The objective of the first lab is to get familiar with the
environment and the Graph class.  It allows to manipulate basic
concepts on graphs.  Neither the algorithms, nor the programming
concepts should be an issue.  The student is asked to write methods
which find the maximum degree of the graph or which construct simple
graphs, complete graphs on $n$ vertices, the complement graph, the
adjacency matrix and the incidence matrix of the graph, etc.

The second series of programming exercises concerns classical search
in graphs: prefix/postfix Depth First Search, Breast First Search. 
Then, the search algorithms are used to work with the connectivity 
of the graph: whether it is connected or not, number of connected 
components. Advanced exercises propose to search for Eulerian and Hamiltonian paths 
and circuits. These search algorithms (DFS and BFS) will be reused in other labs.

The third series of exercises deals with minimum spanning trees.  The
objective is to implement Kruskal's algorithm with the appropriate data
structure, namely a union-find with operations in $\log(n)$.  The
concepts in the exercises are of increasing difficulty.  For instance,
the union-find structure, is first implemented in a table where the
elements indicate the label of the connected component.  If the {\em 
find}
operation is just an access to an element in a table, the {\em union} method
implies a linear number of operations: the whole table has to be scanned
to update the value of the connected component.  Then the student has
to implement the union-find structure as a forest adding step by step
the rank function or path compression. Similarly, for the minimum spanning tree algorithm, first, the very simple inverse Kruskal's
algorithm is asked for. Then, the students must implement Kruskal's 
algorithm with the best union-find structure they could develop. They 
can also implement Prim's algorithm.

On the same principle, a series of programming exercises concerns
Dijkstra's algorithm with various implementations of priority queues
with increasing difficulty.  There is also a lab for a step by step
implementation of Ford and Fulkerson's algorithm for flows.  For less advanced
students, a collection of small labs allows to manipulate simple
combinatorial objects.

\subsection{Feedback on these tools}

Automatic evaluation stimulates students' autonomy since it offers an 
instantaneous and systematic feedback on their work. It allows all 
students to progress at their own pace. The teacher is called only 
for a new learning or when the student is facing a major difficulty. 

Students feel (and tell) that, with this system, the 
professor is more available.  The experience on this tool shows that 
the time the teacher dedicates to the students is of much better 
quality. The tool does all what can be done automatically: gathering 
of student works, syntactic validation and partially functional 
validation. It allows a deep reviewing of the student's work. 
For the teacher also, it offers a support from the automatic 
evaluation system (e.g. results of the automatic evaluation with 
automatic comments, indications of the lines modified by the 
students).

We did not make scientific evaluation of the platform comparing the exam 
results of cohorts using or not the tool mainly because of two reasons. 
The first one is that this tool was very helpful for distant 
learning during the various lockdown periods we encountered and we 
could not deprive some students of the activities of the platform. The other reason is 
that this platform is now widely used  in many courses in the 
universities of the authors and when some professors decide not to use it 
with their groups, we can see that the some of their students  
manage to join the course of the other groups on the platform distorting 
possible analysis.

\section{Advanced shared question banks}
\label{sec:question_bank}

%\com{question partag\'ees / collection de questions similaires 
%structur\'ees / des formats de question avanc\'es}

%\com{et dans la conclusion: plus riche / retenter, \'eviter la 
%triche / apprendre mieux}

In addition to the programming activities, in order to help students
in their progress, questions are proposed for each part of the course.
Those questions can also be part of certificating evaluation in
on-line tests with automated correction.  These are mostly classical
Moodle questions.  The originality here comes from the fact that the
quizzes are shared among a community of teachers through the sharing
space hence creating a very rich collection (see 
Section~\ref{sec:sharing}).  Another originality comes from
the variations of advanced questions based on a predefined non-trivial
framework (Section~\ref{sec:AlgoQuesstions}).  The questions are organized in a bank of questions.  Three
main types exist:

\begin{itemize}
\item basic quizzes that allow students to test themselves their knowledge on vocabulary; definitions, and basic concepts,
\item ``how the algorithms work'' questions, where the students run algorithms on simple graphs;
\item advanced questions in which students have to search more sophisticated answers on graphs.
\end{itemize}

At each step of the course, games and puzzles are proposed to students. These activities want to motivate the students to solve enigmas in a playing context.

To guarantee a large variety of questions, we decided to share the questions proposed by teachers in a bank of questions. 
%This bank contains more than 400 questions \com{(il me semble bien      plus...)}. 
This bank of questions can be used by students for self-assessment, or can be used to develop on-line tests for auto-corrected certificating evaluation.

\subsection{Basic quizzes}
	
\begin{figure}[hptb]
	\center
	\fbox{\includegraphics[width=0.45\textwidth]{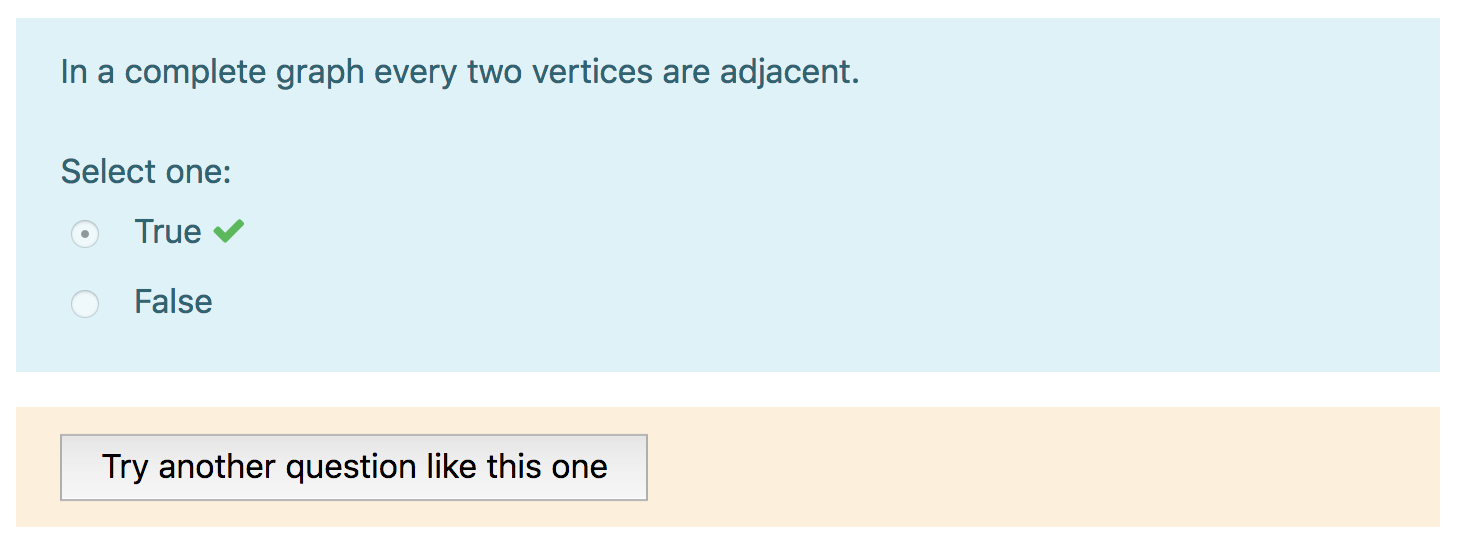}} \\
	\fbox{\includegraphics[width=0.45\textwidth]{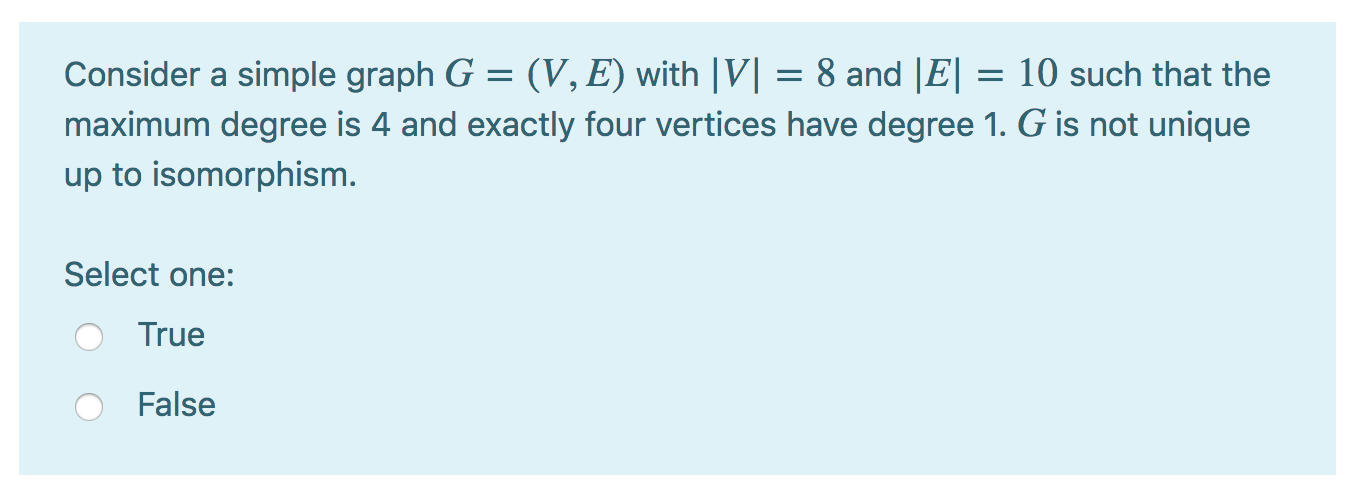}}
	\caption {True-False simple questions selected in the ``Complete''
	and "Degree" categories.} \label{fig:true-false} 
\end{figure}

The basic quizzes are composed of true-false questions and multiple-choice questions, organized in different categories. The more than 400 true-false questions are mainly used to verify that vocabulary and graph definitions are known and understood. Two examples are given, one for the category ``complete graphs'' and one for the category ``Degree'' (see Figure~\ref{fig:true-false}).

\begin{figure}[htb]
\center{\fbox{\includegraphics[width=0.8\textwidth]{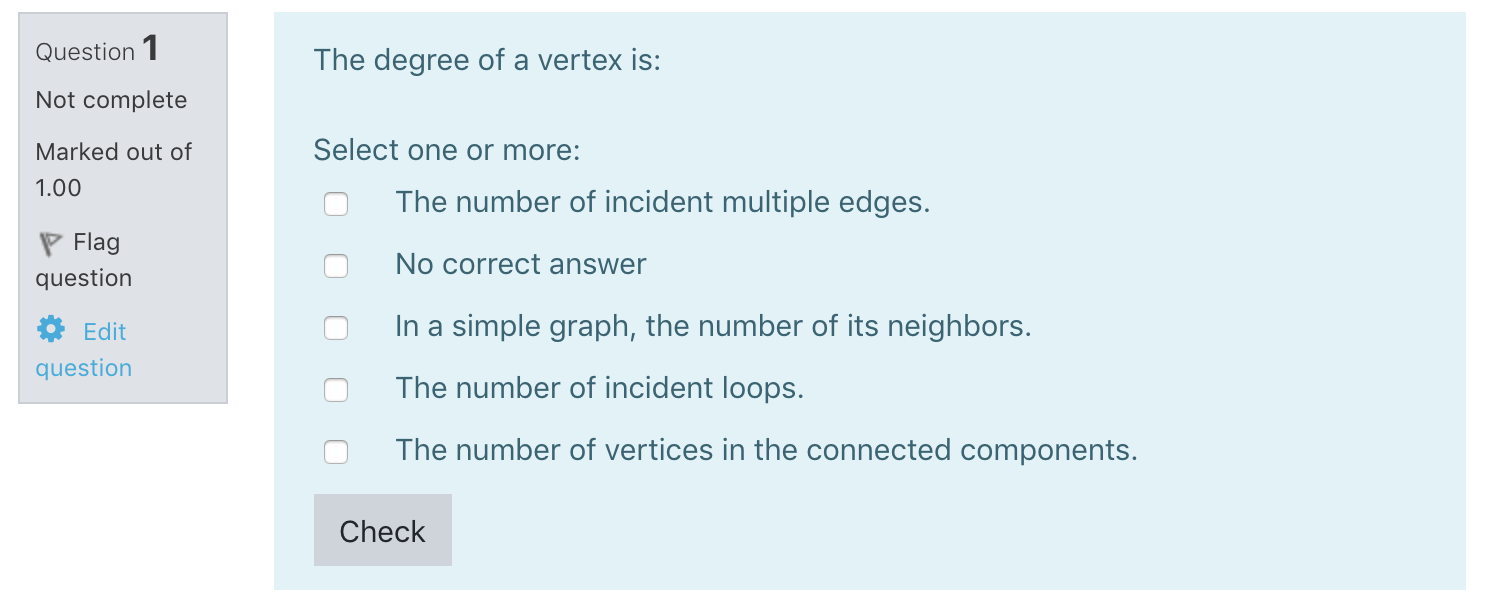}}}
\caption{An example of multiple-choice questions on vertex degree. \label{fig:multiple-choice}}
\end{figure}

Multiple-choice questions are mainly used for more complicated questions, in order to verify if graph concepts are well understood (see for example Figure~\ref{fig:multiple-choice}). 

%\begin{figure}[htb]
%	\center
%	\fbox{\includegraphics[width=0.9\textwidth]{multiple-choice.png}}
%	\caption{An example of multiple-choice questions on vertex degree.}\label{fig:multiple-choice}
%\end{figure}

\subsection{ ``How the algorithms work''  
questions}\label{sec:AlgoQuesstions}

In graph theory courses, classical algorithms are presented. In order to verify that students well understand these algorithms, various activities are proposed. They consist in running the algorithm, step by step, according to the algorithm given in the course. 

\begin{figure}[htb]
\center
{\fbox{\includegraphics[width=0.8\textwidth]{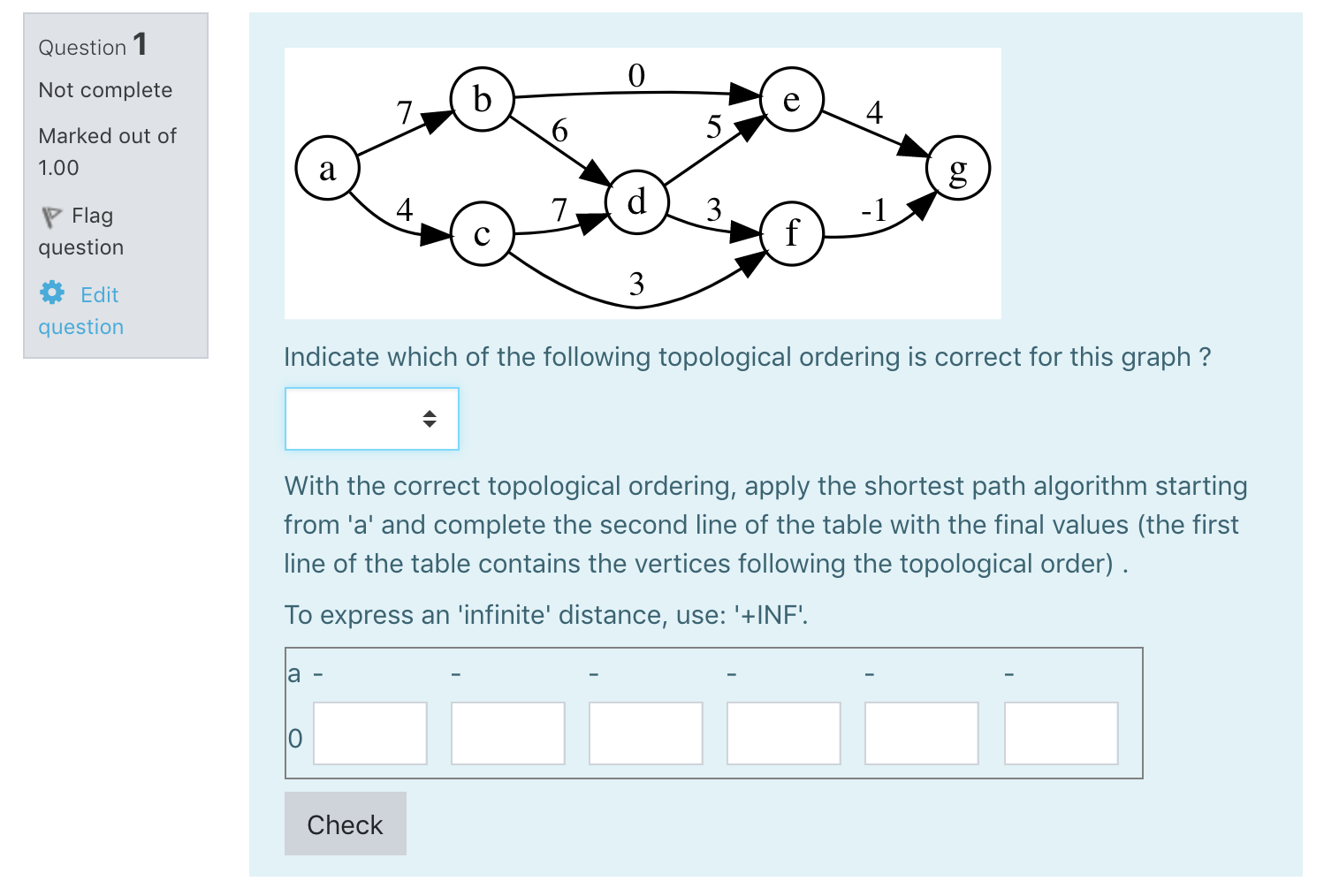}}}
\caption{Shortest path in a Directed Acyclic Graph.}\label{fig:shortest-path-in-DAG}
\end{figure}

Figure~\ref{fig:shortest-path-in-DAG} shows the activity for which the students have to run the shortest path algorithm for Directed Acyclic Graphs. They must first select  a valid topological order from a rolling menu, and then they must give the length of the minimum distance from $a$ to each vertex taken in the chosen topological order.
    
%\begin{figure}[htb]
%	\center
%	\fbox{\includegraphics[width=0.9\textwidth]{shortest-path-in-DAG.png}}
%	\caption{Shortest path in a Directed Acyclic Graph.
%	}\label{fig:shortest-path-in-DAG}
%\end{figure}

\begin{figure}[htb]
\center {\fbox{\includegraphics[width=0.8\textwidth]{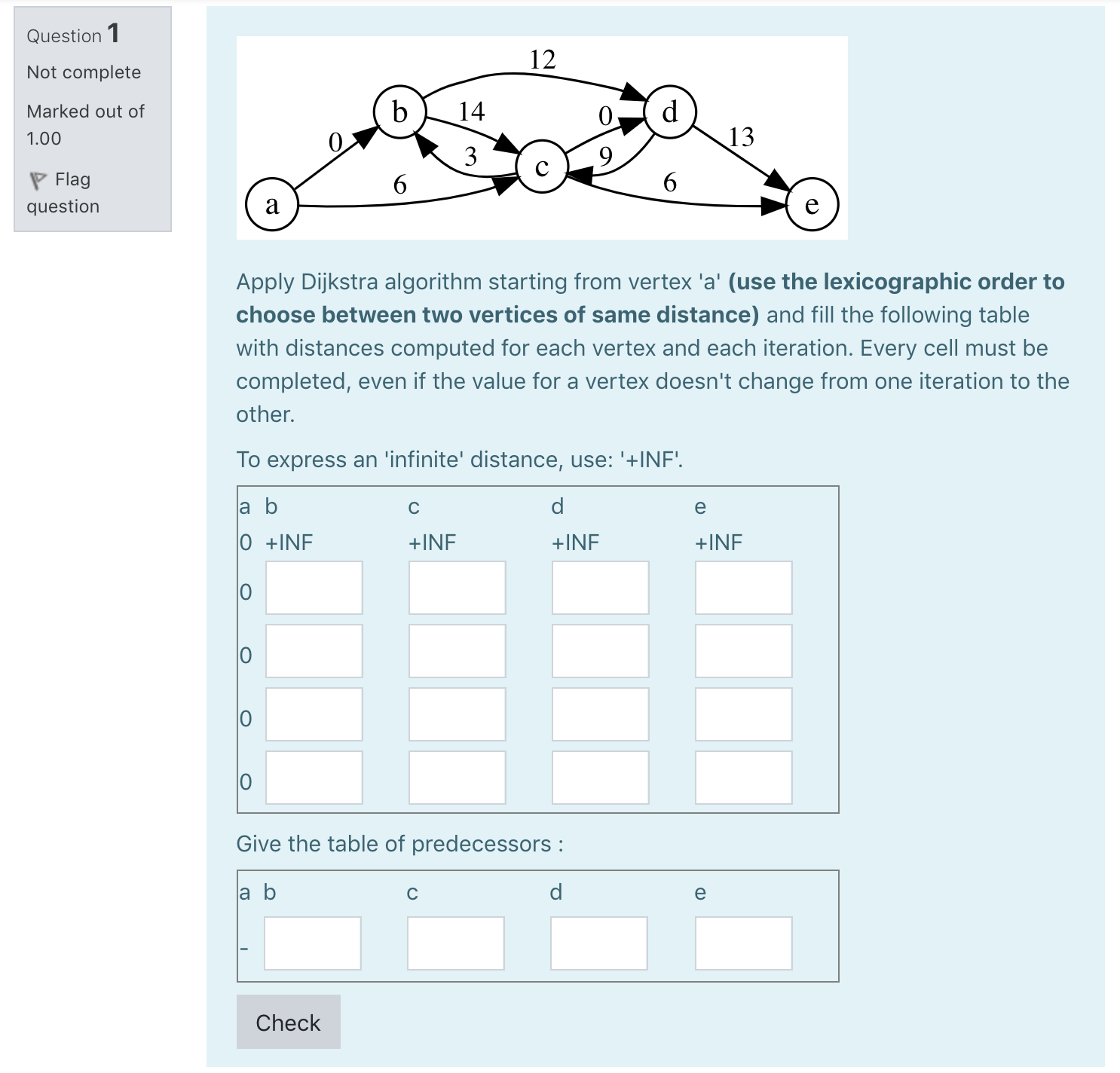}}}
\caption{Dijkstra's algorithm.} \label{fig:shortest-path-dijkstra}
\end{figure}

Similar activities exist for other shortest path algorithms like Bellman-Ford's; An example for Dijkstra's algorithm is given in Figure~\ref{fig:shortest-path-dijkstra}.

%\begin{figure}[htb]
%	\center
%	\fbox{\includegraphics[width=0.9\textwidth]{shortest-path-dijkstra.png}}
%	\caption{Dijkstra's algorithm.
%	}\label{fig:shortest-path-dijkstra}
%\end{figure}

\begin{figure}[htbp]
\center
{\fbox{\includegraphics[width=0.8\textwidth]{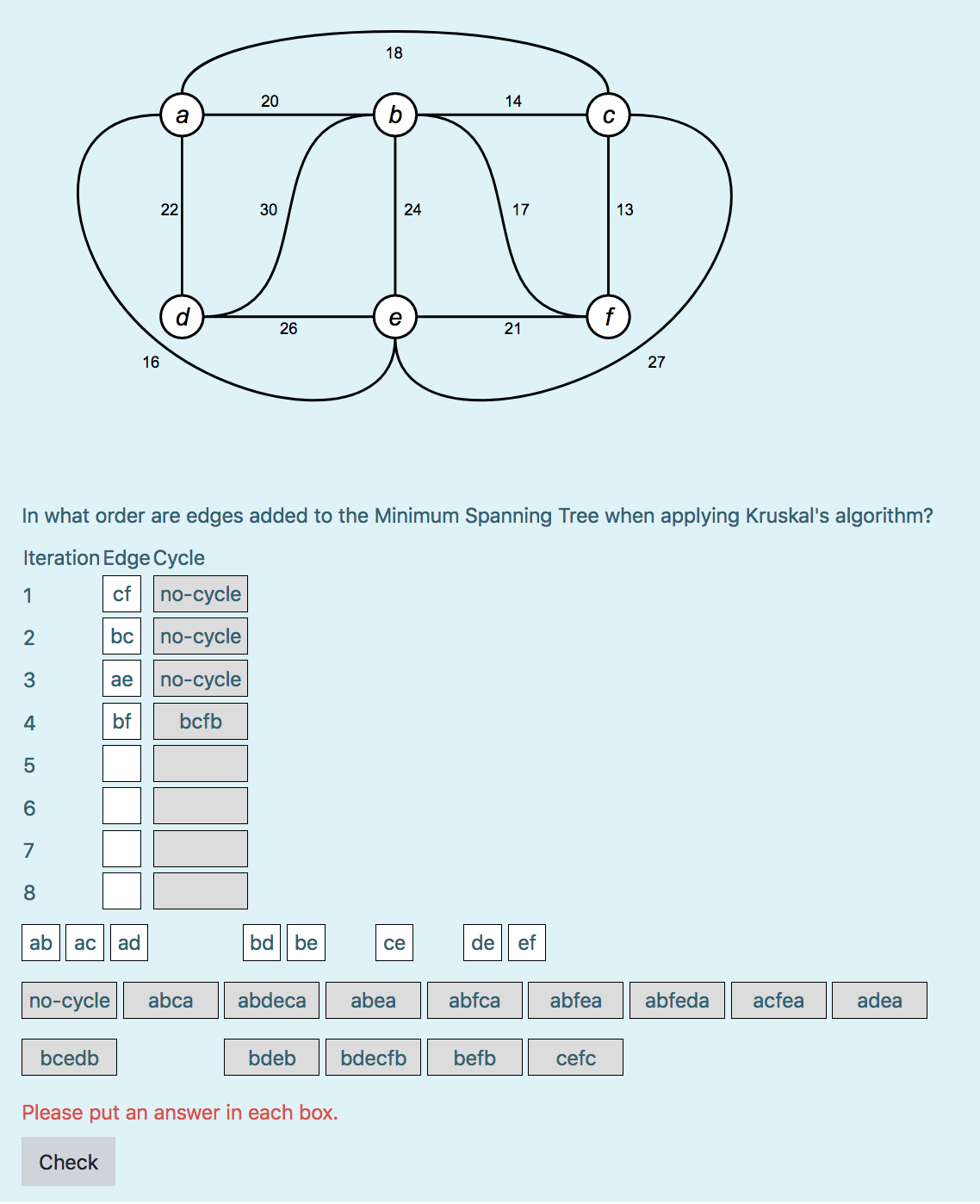}}}
\caption{A complete run of Kruskal's algorithm based on drag and drop questions.} \label{fig:kruskal}
\end{figure}

For the Minimum Spanning Tree, Kruskal's and Prim's algorithm are also proposed.  An example for Kruskal's algorithm is depicted in Figure~\ref{fig:kruskal}.

%\begin{figure}[htb]
%	\center
%	\fbox{\includegraphics[width=0.9\textwidth]{Kruskal.png}}
%	\caption{A complete run of Kruskal's algorithm based on drag and
%	drop questions }\label{fig:kruskal}
%\end{figure}

\begin{figure}[htb]
\center{\fbox{\includegraphics[width=0.8\textwidth]{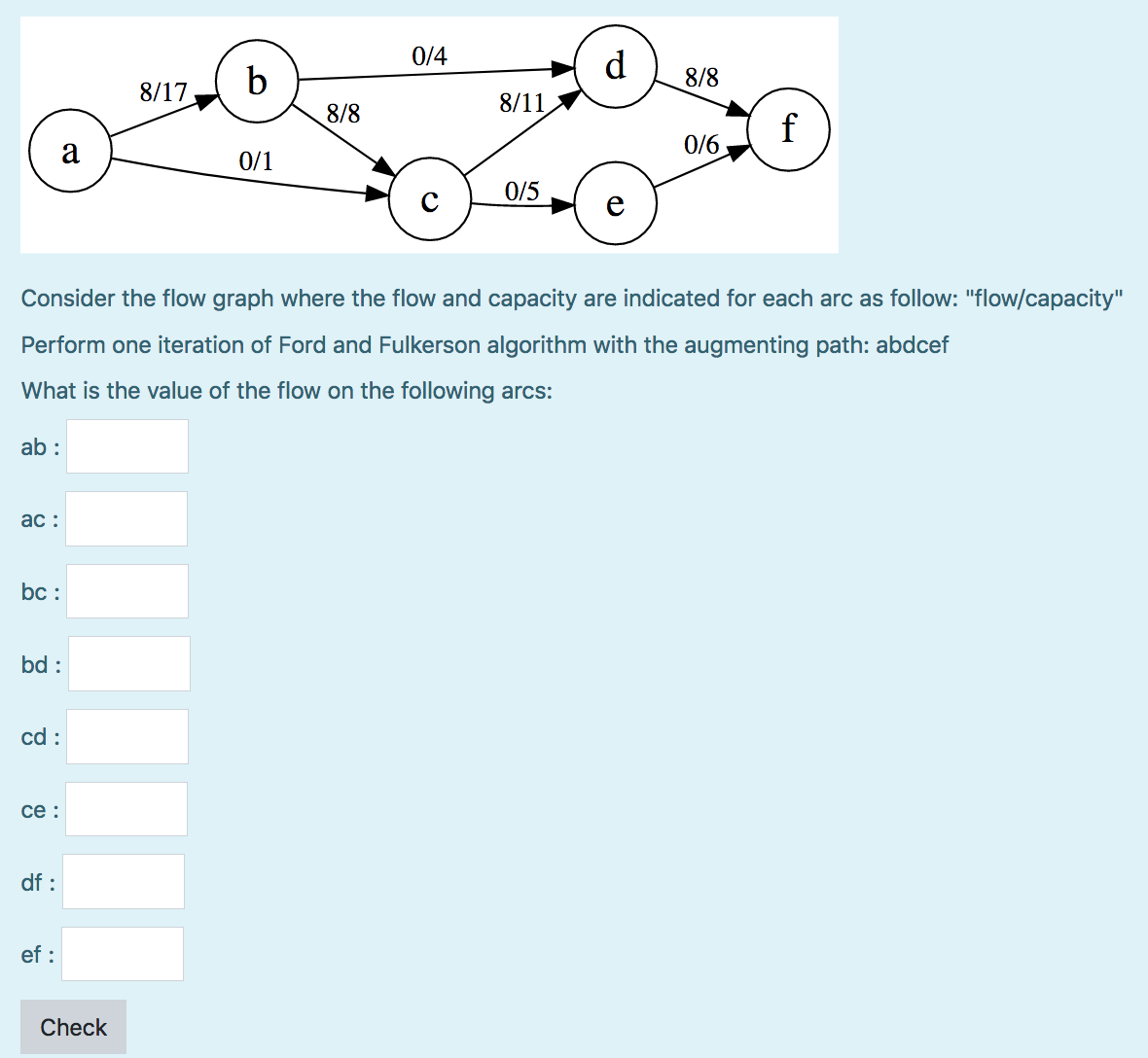}}}
\caption{One iteration of Ford and Fulkerson's algorithm.} \label{fig:flow}
\end{figure}

For the Maximum Flow Problem, Ford and Fulkerson's algorithm has been decomposed into steps; activities on each step have been developed. Some activities ask the students for the validity of a flow in a given network, others ask for augmented paths or the residual graph. An activity asks for the running of one iteration of the algorithm, as shown in Figure~\ref{fig:flow}.  

%\begin{figure}[htb]
%	\center \fbox{\includegraphics[width=0.9\textwidth]{Flow.png}}
%	\caption{One iteration of Ford and Fulkerson's
%	algorithm}\label{fig:flow}
%\end{figure}

For each of the previous described activities, 200 variations have been generated with different graphs or values. This lets students train several times on the same algorithm. This also allows to give different, even though similar, questions to all students.

\subsection{Advanced questions}	

\begin{figure}[htb]
\center
{\fbox{\includegraphics[width=0.45\textwidth]{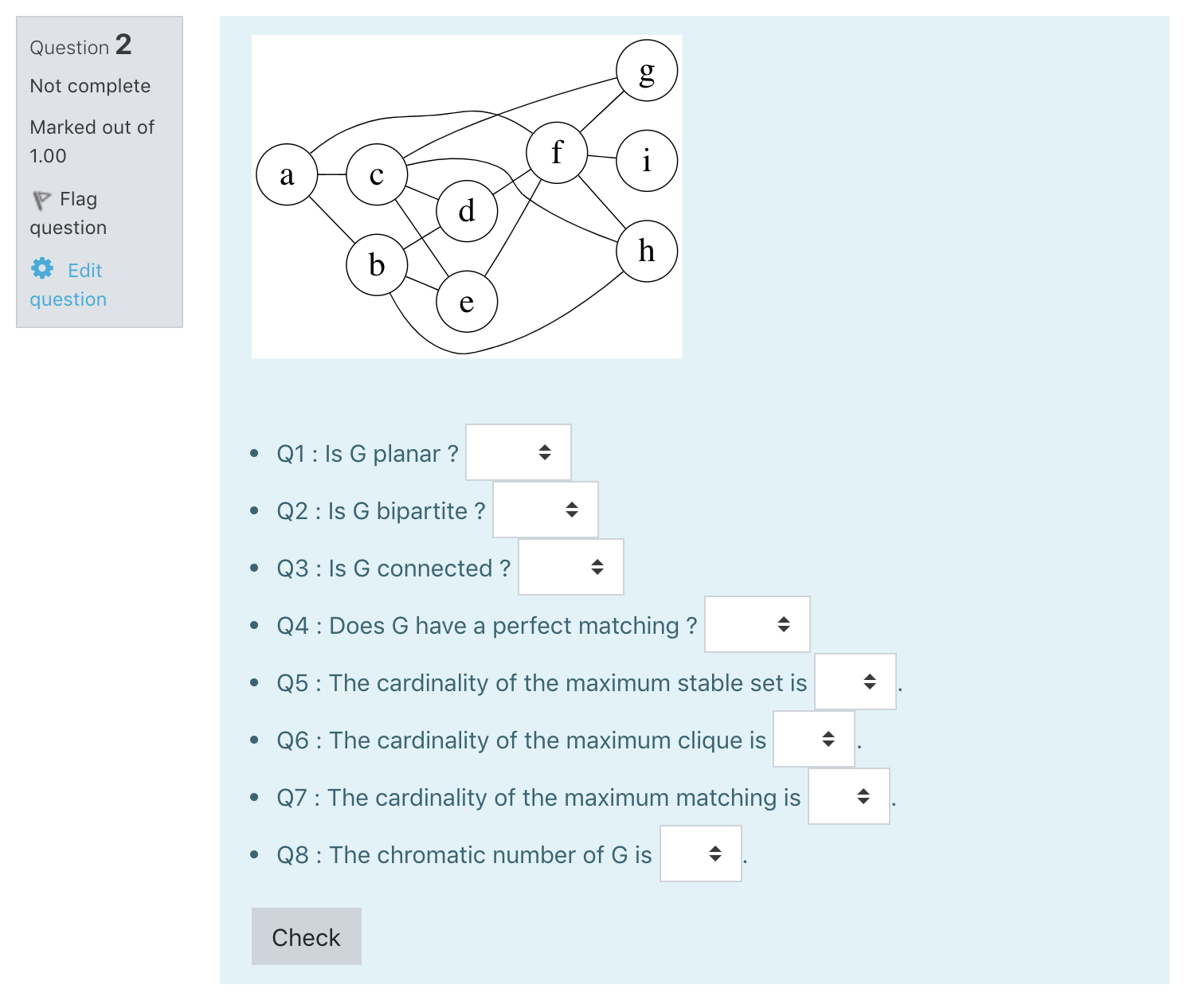}}\ \ {\fbox{\includegraphics[width=0.45\textwidth]{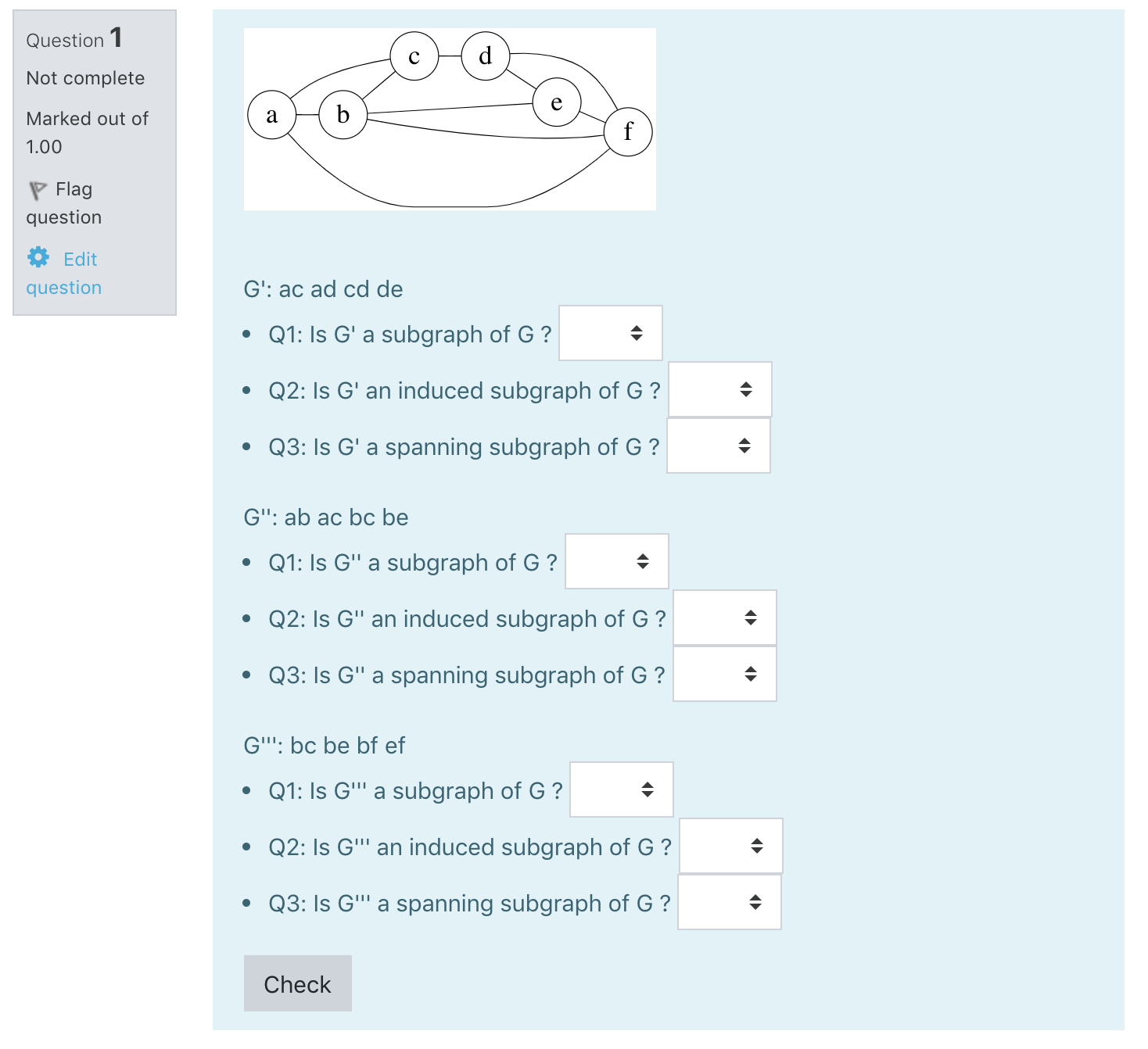}}}}
\caption{An example of questions on structural properties and on subgraphs. }\label{fig:advanced_questions}
\end{figure}

The aim of the advanced questions is to verify that the students perfectly grasped the concepts.
Some questions are focused on structural properties as planarity, connectivity or chromatic number. About 100 different adequate graphs have been automatically generated to set up the pool of questions. For these activities, students have to observe the graph and  work on paper before answering. Some questions ask for a numerical answer that force the student to find arguments that prove its correctness. An example is given in Figure~\ref{fig:advanced_questions}.

%\begin{figure}[htb]
%	\center
%	\fbox{\includegraphics[width=0.5\textwidth]{advanced_questions_structural.png}}
%	\caption{An example of questions on structural properties of graphs.}\label{fig:advanced_questions_structural}
%\end{figure}

Advanced questions have also been generated to train students on the subgraph concept (about 100 different  graphs), as shown in Figure~\ref{fig:advanced_questions}.   

%\begin{figure}[htb]
%	\center
%	\fbox{\includegraphics[width=0.5\textwidth]{advanced_questions_subgraph.png}}
%	\caption{An example of questions on subgraphs of a graph.}\label{fig:advanced_questions_subgraph}
%\end{figure}

%\begin{figure}[htb]
%\FIGURE
%{\fbox{\includegraphics[width=0.5\textwidth]{advanced_questions_subgraph.png}}}
%{An example of questions on subgraphs of a graph. \label{fig:advanced_questions_subgraph}}
%{}
%\end{figure}

\subsection{ Games and puzzles}

\begin{figure}[htb]
\center {
\fbox{%
\includegraphics[width=0.8\textwidth]{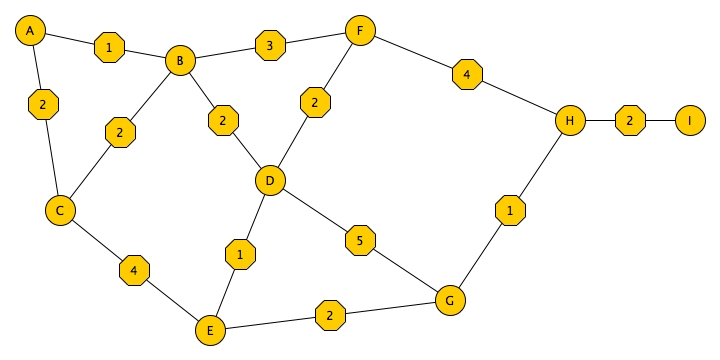}}}
\caption{The riddle example: the map for Anastasia} \label{fig:anastasia}
\end{figure}

The question bank also contains some games for the pleasure because Graphs are well fitted to modeling and solving riddles. We give an example now: 

{\em Most of the time, Anastasia is a physiotherapist in Liège, Belgium. During the summer (July and August), she works in the Alps as a mountain guide on the great hiking trails. She leads groups of hikers along trails, spending one night in a refuge with each group. She operates as follows. She picks up a group at one station, leads them to a refuge for an overnight stay, and then leads them down to another station the next day. There, she either continues on with the same group, or takes on a new group. She has an agreement with these refuges: at the end of the season, each refuge sends her an invoice for the number of nights she spent there. In order to save money, she only sleeps in refuges (as a guide, she gets a discount in refuges).

Anastasia is not very well organized. At the end of September, she remembers that she has to pay the refuge bills. She got back to her physiotherapy life and the Alpine trails feel very far away. She can no longer recall the exact sequence of hikes, where she started, or the last hike she did. All she can remember is arriving in France by train, hitchhiking to a mountain station, and returning in the same manner. Fortunately, thanks to the train tickets that she found, she can work out the total number of nights she spent in refuges.

However, after adding up the nights on all the invoices, there is one too many. Did a refuge mistakenly charge her an extra night?

The map on Figure~\ref{fig:anastasia} shows the possible hikes in the 
area where Anastasia was and the number of nights indicated on the 
invoice of each refuge. The letters are the stations and the octagons the refuges. Looking at this figure, can you find the refuge with the wrong number of nights?\footnote{The reader certainly recognized a eulerian path problem. The graph has 4 odd vertices but only two of them are adjacent!}} 

\subsection{United we stand}	

Creating technologically advanced questions is time consuming.
Therefore, the idea is to share them within a community.  This also
improves the quality and visibility of the exercises since sharing a
content implies a benevolent reviewing from the pairs.  Having a very
large bank of similar questions allows multiple training but also 
avoids cheating and copying among students.  Technically, we use the
question bank tool which is available in the Moodle LMS. A mediator
organizes the question bank adding comments and the questions are
presented in context in the open course to help appropriation by the
community.

\section{Caseine platform}
\label{sec:caseine}

%\com{Bernard}

%\com{Structure de cette section: Architecture, contenu, partage, benefices}

%\com{\'evaluation des \'etudiants / facili\'e pour monter des cours, 
%utiliser des supports existants}

%Section~\ref{sec:autres} compares the caseine platform to existing similar
%projects.  
Sections~\ref{sec:AutoLP} and~\ref{sec:question_bank}
presented innovative tools to enrich courses on graph theory.
Thousands of questions have been generated, from simple true-false
questions to more sophisticated ones.  Numerous and various
programming labs have also been developed.  All these activities are
available for students and teacher on a unique open platform, {\tt
caseine.org}, based on Moodle.

%There are about 400 questions
%in this category, organized by subject (see
%Figure~\ref{fig:bank}).
%
%	\center
%	\fbox{\includegraphics[width=0.45\textwidth]{Bank.png}}
%	\caption{The True-False question bank.  The numbers between
%	parenthesis indicate the number of questions in each
%	category}\label{fig:bank}
%\end{figure}

\subsection{A learning platform with content sharing and code evaluation}\label{sec:autres}

The caseine.org platform relies on an open and participative project
which gathers teachers not only from worldwide universities but also
from high-school or from associations or companies with collective
interest.  For instance, it is used by a 
company\footnote{\url{https://le-campus-numerique.fr/}} training people for who
it is difficult to enter directly a classical academic training (e.g.
persons with autism).  

The caseine.org platform is based on Moodle~\cite{Moodle}, a widely used Learning Management
System (LMS), which provides students with a learning environment and let teachers monitor students' progress. It is aimed simultaneously at teachers who create courses and content and monitor students' progress, at institutions which can manage groups, and at students for their training. Compared to a classic instance of Moodle, caseine is original in that it offers a single environment where, on the one hand, students can benefit from an automatic evaluation of their mathematical models and
programming codes\footnote{using the VPL plugin \cite{vpl}} and on the other hand, teachers can create self-assessed activities, share them with the
community and use shared activities. 

While other learning platforms exist,  we do not know of any open 
platform  offering at the same time (1) a pedagogical environment for 
university training where the teacher can build their own course and 
(2) coding exercises with automatic and self-evaluation and (3) all 
sorts of pedagogical activities shared by an open community of 
teachers. Other platforms only address one or two of these three 
aspects (see \cite{Cambazard21} for some examples of such platforms).

Teachers from various universities\footnote{In Septembre 2022, 14 
universities  and 12 high-schools officially indicate that they use the platform.} in the world use the platform with 
their students. Indeed, caseine is an open academic tool, free for 
non-commercial use, and any teacher or student can straightforwardly 
connect to the platform with their own academic login, as long as 
their university is a member of the worldwide Edugain network 
\cite{edugain}. Otherwise, students or teachers can request a manual 
account. The platform also offers open courses in Graph theory were 
anybody with a 
mathematical or computer science background can connect and start 
learning. This opportunity was used in France by many high-school 
teachers in France when graph theory was introduced in the national 
program for high-school students.

During academic year 2020-2021, the platform was used by more than 
10.000 users: up to 1500 users per day were connected during the 
COVID-19 lockdown periods in France, since March 2020. 

\subsection{Organization}

%The architecture of the platform is shown Figure~\ref{fig:caseine_architecture}. 
%\com{il faudrait ajouter quelques details techniques: interface web, BDD, etc. pour que le lecteur comprenne bien l'objet. Mais je ne connais pas assez les caracteristiques techniques pour le faire bien. Mais je veux bien apprendre~!}

%\begin{figure}[htb]
%	\center
%	\fbox{\includegraphics[width=0.9\textwidth]{todo_architecture.png}}
%	\caption{\com{Caseine architecture (todo)}.}\label{fig:caseine_architecture}
%\end{figure}

%\begin{figure}[htb]
%\FIGURE
%{\fbox{\includegraphics[width=0.9\textwidth]{todo_architecture.png}}}
%{\com{Caseine architecture (todo).} \label{fig:caseine_architecture}}
%{}
%\end{figure}

%\com{Contenu, originalite de la plateforme et ce qu'on y trouve}
 
As on a classical Moodle platform, each teacher scripts courses 
composed of activities. On caseine, they can also insert in the course activities that other teachers decided to share to the community. And of course, they can decide, for each of their own activities, whether to share it or not. The students have access to the course thus created, and to all the resources that their teacher has made available. 
 
The originality of caseine derives from the fact that, on the same platform, on the one hand, mathematical models and programming codes can be evaluated automatically and in autonomy and, on the other hand, teachers can create auto-evaluated activities, share them to the community and use shared activities in their own courses.

The activities and exercises described in this paper can be tested in
the Graph Open Course on the platform: \url{https://caseine.org/course/view.php?name=GraphsOpen}.  The course is
partly in French but the quizzes are in English.  A page dedicated to
this paper with examples of bilingual activities is: \url{https://caseine.org/mod/page/view.php?id=25138}.  Beyond Graph theory,
caseine offers courses on general OR tools like Linear Programming,
Mixed Integer Programming, Dynamic Programming, Constraint
Programming...  It also contains specialized courses like production
planning and logistics and practical case studies.  Apart from OR,
caseine is used in various courses in Computer Science, Industrial
Engineering and Mathematics from secondary school to master programs.
Those courses also use original automatic tools and the activity
sharing tools presented in the context of graphs.

\subsection{The sharing space and the visitable courses}\label{sec:sharing}
\begin{figure}[htb]
\center{
\fbox{%
\includegraphics[width=0.8\textwidth]{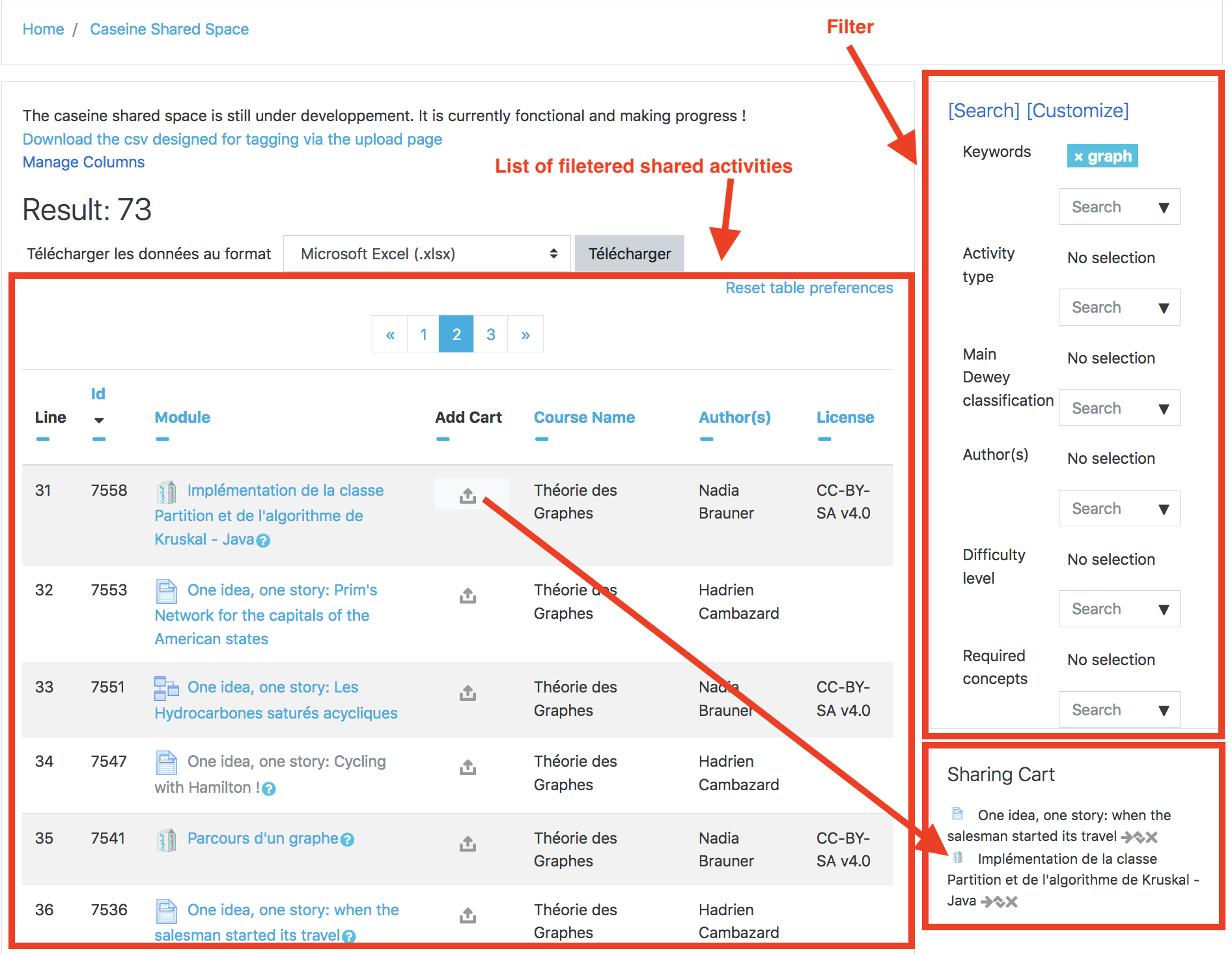}}}
\caption{The search space lists the shared activities of the platform 
and the {\em sharing cart} allows to choose the activities that will be inserted in the user course} \label{fig:sharingSpace}
\end{figure}

Caseine was first developed with the main objective to share contents 
within a teacher community. Even if some contents can be private, an 
increasing part of the contents is now available for the community. 
Teachers can freely use the shared contents to create a new course or 
to improve an existing one, and create others that they may or may 
not share. They can thus propose their own exercises to the teaching community through the sharing space under licenses that allow sharing (free license, Creative Commons-BY-SA...).
Adding an activity  to a course  from the sharing space is like adding
any type of activity.  The search space allows to filter the 
activities
based on chosen metadata and insert them into the course just with
one click.  A sharing cart also allows to choose a set of activities (see on Figure~\ref{fig:sharingSpace} the search space with the filter and the sharing cart).

\begin{figure}[htb]
\center{
\fbox{%
\includegraphics[width=0.6\textwidth]{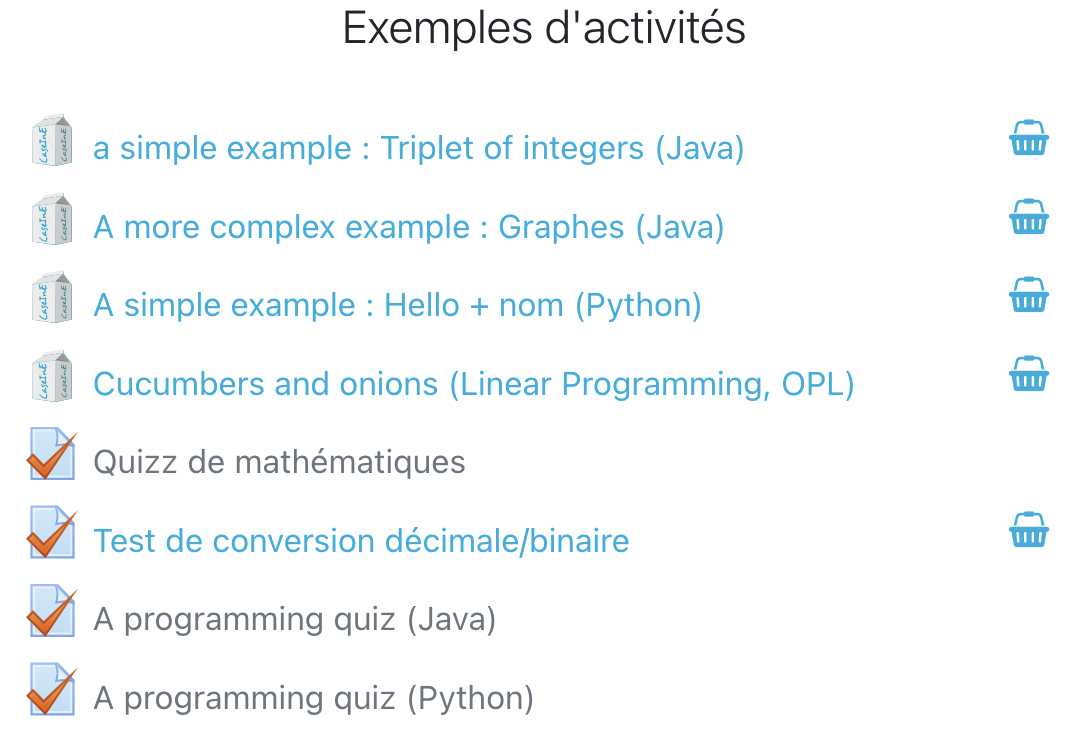}}}
\caption{Example of a visitable course section: Shared activities (in 
blue) can be added to the visitor's cart who can then insert them in 
their course.
} \label{fig:visitableCourse}
\end{figure}

%{\color{red} 
A teacher cans also decide to make their course 
visitable, i.e. to open a partial access to the teachers of the 
community so that they can consult (and retrieve) the shared 
activities directly in their context (see 
Figure~\ref{fig:visitableCourse}). The visitors can then add the shared 
activities to their cart.
%}

The terms and conditions are detailed on the platform. The basic 
idea is that the platform is free for people not doing money with
it. Contribution to the maintenance's cost of the platform can be asked
for, otherwise. Self-training and visiting in open courses is free.

\subsection{Better learning - better teaching}

Since the launch of the platform and the sharing of contents, the benefits for teachers and students have been significant.
% After a few years of use, the benefits brought by Caseine are numerous.

Digital tools can make a significant contribution to teaching, particularly in the teaching of mathematics and computer science. These tools make the student active and autonomous in a large number of activities. 
For students, caseine is a unique portal that brings together course materials, various self-assessment activities, and evaluated tests. The modeling and programming activities allow them to practice, and to have direct feedback from the teacher in their programs (see section~\ref{sec:AutoLP}). Thanks to the platform, the students can follow their progress, know their results in the evaluation tests, and thus see if they are at the expected level.
The presented teaching tools focus the student active role by increasing
engagement and autonomy of students. The proposed activities better fit various types of students' needs: advanced students can find activities that allow to go further and deeper in the knowledge and slower students can train as many times as they need. On the platform there is a
large variety of uses of those tools from complete autonomy
(personal/team work, at home/during the course) to traditional
classroom with validation in autonomy or as a support to conduct a 
flipped classroom. Finally, a teacher can provide direct feedback on 
a program or mathematical model written by a student by annotating 
the content. We believe that such feedback can have a key effect on 
the learning as outlined by John Hattie \cite{Hattie09}:
"In summary, feedback is one of the most powerful influences on 
learning."

%\com{ref sur bénéfice apporté aux étudiants par un retour rapide précis et de qualité ?
%J’ai bien la phrase de Dehaene mais elle est en francais et pas dans un article scientifique. "[…] Le retour sur erreur est le troisième pilier de l’apprentissage, et l’un des paramètres éducatifs les plus influents: la qualité et la précision du retour que nous recevons déterminent la rapidité avec laquelle nous apprenons."
%Stanislas Dehaene, Apprendre !, ed. Odile Jacob}

For the teachers, the platform allows them to script their courses, to propose various activities to the students. It allows the teachers to follow the progress of each student, and to guide them. For programming, they have access to the student's program, they can execute it and propose corrections or improvements directly in the code. They can also set up automatic evaluation tests. For automatic evaluation, the teacher team can and must come up with new evaluation methods better fitted to new technological tools, to expectation and needs of students and complementary to classical evaluation methods. This issue being new, it takes time and requires new ideas and developments.  This is the reason why it is built by a community sharing ideas and contents. Moreover, collaboration between teachers  allows both to improve the content offered and to better understand the learning difficulties of different audiences.

\section{Conclusion and Future work}

In this paper, we have presented programming activities and quizzes. 
Other types of course material are shared on the platform like the {\em One idea, one story} pages which put the results in a historical 
perspective or a shared databases of graph drawings in 
Tikz\footnote{\url{https://moodle.caseine.org/mod/data/view.php?id=21600}}.

In graph theory, one important perspective is to be able to 
settle/evaluate the complexity of the algorithms proposed by the 
students without downgrading the response time of the system. 
Preliminary results in this direction are encouraging. 

If you might be interested in using these tools or joining the 
adventure, you can start by visiting the site caseine.org.

% Acknowledgments here
{\bf Acknowledgment} {The authors would like to thank the two engineers of the platform, Astor Bizard and Florence Thiard who develop the tools and help teachers use the platform. 

This work has been partially supported by the Idex Université Grenoble-Alpes (ANR-15-IDEX-0002) and by the LabEx PERSYVAL-Lab (ANR-11-LABX-0025-01), both funded by the French program Investissement d’avenir.
%
% Enter the text of acknowledgments here
}% Leave this (end of acknowledgment)

% Appendix here
% Options are (1) APPENDIX (with or without general title) or 
%             (2) APPENDICES (if it has more than one unrelated sections)
% Outcomment the appropriate case if necessary
%
% \begin{APPENDIX}{<Title of the Appendix>}
% \end{APPENDIX}
%
%   or 
%
% \begin{APPENDICES}
% \section{<Title of Section A>}
% \section{<Title of Section B>}
% etc
% \end{APPENDICES}

% References here (outcomment the appropriate case) 

% CASE 1: BiBTeX used to constantly update the references 
%   (while the paper is being written).
\bibliographystyle{plain} % outcomment this and next line in Case 1
%\bibliography{<your bib file(s)>} % if more than one, comma separated

% CASE 2: BiBTeX used to generate mypaper.bbl (to be further fine tuned)
%\input{mypaper.bbl} % outcomment this line in Case 2

\end{document}